\documentclass[12pt,a4paper]{article}

\usepackage{amsmath,amssymb}
\allowdisplaybreaks[4]

\begin{document}

\bibliographystyle{unsrt}

\begin{titlepage}

\title{\bf Function group approach to unconstrained Hamiltonian
Yang--Mills theory}
\author{Antti Salmela\footnote{Email: Antti.Salmela@Helsinki.Fi} \\
\it Theoretical Physics Division \\ \it Department of Physical Sciences \\
\it P.O. Box 64, 00014 University of Helsinki, Finland}
\date{}

\maketitle

\thispagestyle{empty}

\begin{abstract}

Starting from the temporal gauge Hamiltonian for classical pure
Yang--Mills theory with the gauge group SU(2) a canonical
transformation is initiated by parametrising the Gauss law
generators with three new canonical variables. The construction of
the remaining variables of the new set proceeds through a number
of intermediate variables in several steps, which are suggested by
the Poisson bracket relations and the gauge transformation
properties of these variables. The unconstrained Hamiltonian is
obtained from the original one by expressing it in the new
variables and then setting the Gauss law generators to zero. This
Hamiltonian turns out to be local and it decomposes into a finite
Laurent series in powers of the coupling constant.

\end{abstract}

{\noindent PACS: 11.15.Kc, 11.10.Ef, 02.20.Sv}

\end{titlepage}

\pagebreak

\newenvironment{acknowledgements}{\hfill \vspace{12pt plus 5pt minus
3pt} \break \bf Acknowledgements \rm \hfill \vspace{8pt plus 3pt minus
1.5pt} \break}{} 
\newcommand{\calPi}{\text{\Large $\pi$}}
\newcommand{\calA}{\mathcal{A}}

\section{Introduction}

An important and still open problem of quantum chromodynamics is to
work out analytical predictions for the low-energy states of the
theory. In order to make these predictions we need a proper quantum
Yang--Mills theory which is valid in the low-energy regime. However,
for many reasons it has turned out to be a difficult task to construct
a useful physical Hamiltonian. One of the problems encountered is the
implementation of Gauss's law in the Hamiltonian formalism. Up to this
date, several methods have been developed to tackle it \cite{gj} --
\cite{kp}, and this paper aims to provide a novel method, which is
motivated by Lie's theory of function groups and their canonical
representations.

Usually one starts with an extended quantum Hamiltonian where the
physical subspace consists of states that are annihilated by the Gauss
law generators. In this paper, by contrast, the order of quantisation
and constraining is reversed and Gauss's law is incorporated into the
Hamiltonian formalism already at the classical level with the help of
a suitable canonical transformation. Whenever one performs canonical
transformations in a classical Hamiltonian gauge theory, one must
choose the new variables in a way that makes their fundamental Poisson
bracket relations compatible with the gauge algebra satisfied by the
Gauss law generators. This is often done by the method of
Abelianisation, where the Gauss law generators are multiplied by
suitable matrices that transform them into mutually involutive
canonical momenta. In this paper, however, the opposite strategy is
followed and the generator algebra is taken as given. The generators
are then parametrised with the minimum number of canonical variables
in such a way that the gauge algebra is satisfied as a consequence of
the fundamental Poisson brackets of the new variables. The remaining
variables of the new set are finally constructed by following the
logical steps implied by this parametrisation. The procedure is
carried through for pure SU(2) Yang--Mills theory, but a
generalisation to other Lie groups is discussed in the end.

The actual construction of the canonical transformation is done in
several steps in section \ref{konstru}. The procedure is a bit
lengthy, but I prefer to give a presentation where the underlying
logic is made clear and where possibilities for modifications and
generalisations are also offered. The final transformation is then
used in the third section, where the unconstrained Hamiltonian is
derived and expanded in a finite series involving both positive and
negative powers of the coupling constant. The last section is devoted
to conclusions. Throughout the paper I will use Einstein's summation
convention with spatial and Lie algebra metrics normalised to positive
unity. The generators of the SU(2) algebra are, as usual, taken to be
$T_a = \frac{1}{2} \sigma_a$, where the $\sigma_a$'s stand for the
Pauli matrices.

\section{Construction of the canonical transformation}
\label{konstru}

\subsection{Parametrisation of the Gauss law generators}
\label{gpar}

We start with the temporal gauge ($A_0^a = 0$) Hamiltonian
\begin{equation} \label{ham}
H = \int \! \left( \frac{1}{2} \, \Pi_{ka} \Pi^{ka} + \frac{1}{4}
\, F_{kl}^a \, F^{kl}_a \right) d^3 {\bf x},
\end{equation}
where the field tensor $F_{kl}^a$ is defined by
$$ F_{kl}^a = \partial_l A_k^a - \partial_k A_l^a + g \,
{\varepsilon_{bc}}^a A_k^b \, A_l^c. $$
The variables $A_k^a({\bf x})$ and $\Pi_{ka}({\bf x})$ are
canonically conjugate, i.e., they satisfy the fundamental Poisson
bracket relations
$$ \{ A_k^a({\bf x}), \Pi_{lb}({ \bf y}) \} = \delta_{kl} \,
{\delta^a}_b \, \delta({\bf x} - {\bf y}). $$ 
From these relations it follows that the Gauss law generators
\begin{equation} \label{gdef}
G_a = \partial^{\, k} \, \Pi_{ka} - g \, {{\varepsilon_b}^c}_a
A^{kb} \, \Pi_{kc}
\end{equation}
obey the SU(2) algebra
\begin{equation} \label{alg}
\{ G_a({\bf x}), G_b({\bf y}) \} = - g \, {\varepsilon_{ab}}^c \,
G_c({\bf y}) \delta({\bf x} - {\bf y}).
\end{equation}
They also generate time-independent gauge transformations of the
canonical variables as follows:
\begin{eqnarray} \label{transom}
\{ G_a({\bf x}), A_k^b({\bf y}) \} &=& - {\delta_a}^b \,
\partial_k^{(x)} \delta({\bf x} - {\bf y}) - g \,
{\varepsilon_{ca}}^b \, A_k^c({\bf y}) \, \delta({\bf x} - {\bf
y}), \nonumber \\* \{ G_a({\bf x}), \Pi_{kb}({\bf y}) \} &=& - g
\, {\varepsilon^c}_{ab} \, \Pi_{kc}({\bf y}) \, \delta({\bf x} -
{\bf y}).
\end{eqnarray}
The canonical equations of motion
\begin{equation} \label{cane}
\dot{A}_k^a({\bf x}) = \frac{\delta H}{\delta \Pi^k_a({\bf x})},
\hspace{2em} \dot{\Pi}_{ka}({\bf x}) = \, -\, \frac{\delta H}{\delta
A^{ka}({\bf x})}
\end{equation}
reproduce the dynamical Yang--Mills equations
$$ \ddot{A}_k^a({\bf x}) - \left[ {\delta^a}_c \, \partial^{l} - g \,
{\varepsilon_{bc}}^a A^{lb}({\bf x}) \right] F_{kl}^c({\bf x}) = 0, $$
but not Gauss's law
$$ G_a({\bf x}) = 0. $$
However, the Gauss law generators are constants of motion, i.e.,
$$ \dot{G}_a({\bf x}) = 0 $$
in the dynamics described by the equations (\ref{cane}). This property
ensures that the implementation of Gauss's law can be done
consistently with the Hamiltonian equations of motion.  Unfortunately
we cannot just use equation (\ref{gdef}) to eliminate redundant
coordinates in the limit $G_a \rightarrow 0$, because we do not know
which coordinates to eliminate or how to deal with the canonical
conjugates of these redundant variables.

The first stage in the function group approach consists of replacing
the Gauss law generators with such canonical variables that will
vanish in the limit when Gauss's law is put into force.  At this point
we recall that in Lie's work a function group is defined as a set of
variables equipped with Poisson brackets that close on the set
\cite{lie}. According to Lie, every function group can be transformed
into a form where every variable either has a canonically conjugate
counterpart in the set or its Poisson brackets with the remaining
variables vanish. Applying this idea to the function group formed by
the $G_a$'s, we parametrise it with three canonical variables $p_1$,
$p_2$ and $q_2$ as follows:
\begin{eqnarray} \label{trans}
G_1 &=& \sqrt{p_1^2 - p_2^2} \, \cos (g \, q_2) \nonumber \\*
G_2 &=& -\sqrt{p_1^2
-p_2^2} \, \sin (g \, q_2) \\*[1.2ex]
G_3 &=& p_2. \nonumber
\end{eqnarray}
It is easy to check that the SU(2) algebra relations (\ref{alg}) are
satisfied if the Poisson brackets of the new variables are canonical,
i.e., if
$$ \{ q_2({\bf x}), p_2({\bf y}) \} = \delta({\bf x} - {\bf y}) $$
and all the other brackets vanish. Conversely, we can invert this
transformation and check that the variables
\begin{eqnarray} \label{intrans}
p_1 &=& \sqrt{G_1^2 + G_2^2 + G_3^2} \nonumber \\*[1.8ex] p_2 &=&
G_3 \\* q_2 &=& - \, \frac{2}{g} \arctan \left( \frac{\sqrt{G_1^2
+ G_2^2} - G_1}{G_2} \right) \nonumber
\end{eqnarray}
satisfy the fundamental Poisson bracket relations by virtue of the
algebra (\ref{alg}).

The parametrisation (\ref{trans}) is by no means the only possibility
of defining a canonical representation, but it is one of the simplest
with respect to the properties of the SU(2) algebra. Namely, equation
(\ref{alg}) allows us to identify the $G_a$'s with the basis vectors
of the SU(2) algebra and the Poisson bracket with the commutator. We
can now make use of the fact that for all semisimple Lie groups the
Casimir operators together with the basis of the Cartan subalgebra
span an Abelian subspace of the enveloping algebra. With
higher-dimensional Lie groups this Abelian subspace can be augmented
by Casimir operators of some lower-dimensional subalgebras. Since all
canonical momenta must have vanishing Poisson brackets with each
other, we see that the maximal set of momenta can be obtained from the
maximal Abelian subspace of the enveloping algebra. This is the idea
behind the transformation (\ref{intrans}), where we now recognise
$p_1^2$ as the Casimir operator of SU(2) and $p_2$ as the usual basis
vector for the Cartan subalgebra. Once this choice has been made, the
form of $q_2$ follows from the consistency of the canonical Poisson
brackets with the algebra (\ref{alg}). However, the fundamental
Poisson bracket relations do not determine the canonical conjugate of
$p_1$ uniquely and we can thus leave the specific form of $q_1$ open
at this stage. The next step is to extend the transformation
(\ref{intrans}) to the remaining variables.

\subsection{Gauge-invariant variables} \label{givv}

Let $\xi_i$ stand for any new canonical variable not equal to those
already fixed. By the requirement that the Poisson brackets between
$\xi_i$ and the members of the set $\{q_1,p_1,q_2,p_2\}$ vanish and
with the help of the parametrisation (\ref{trans}) we see that
\begin{subequations} \label{ginv}
\begin{eqnarray}
&& \{ G_a({\bf x}), \xi_i({\bf y}) \} = 0, \qquad a = 1, 2, 3,
\label{ginv1} \\*[1.2ex]
&& \frac{\delta \xi_i({\bf y})}{\delta p_1({\bf x})} =
0. \label{ginv2} 
\end{eqnarray}
\end{subequations}
In particular, equation (\ref{ginv1}) means that all the remaining
variables must be invariant under topologically trivial gauge
transformations. Since we have already defined three
non-gauge-invariant variables ($q_1, q_2, p_2$), they completely fix
the gauge angles (modulo constant gauge transformations) in the new
set of variables. The gauge-invariant fields must therefore be
constructed by transforming the old variables into a gauge where
$q_1$, $q_2$ and $p_2$ are absent. Note that the term "gauge" does not
imply neglecting any dynamical degrees of freedom at this stage, it
only describes the way that the gauge-invariant variables of the new
set are formed. In other words, the new variables consist of both
gauge-dependent and gauge-invariant degrees of freedom. Although it
may sound a little paradoxical, the gauge-invariant variables also
satisfy a gauge condition due to the fact that, by construction, the
gauge-dependent degrees of freedom have been transformed away. (This
procedure is discussed in a more general context in Ref. \cite{cmt}.)

Let us begin with the elimination of $q_2$ and $p_2$. When these
variables tend to zero, equation (\ref{trans}) shows that the
components $G_a$ tend to $\delta_{a1} \, p_1$. The intermediate
variables $\widehat{A}_k^a$ and $\widehat{\Pi}_{ka}$ are then
determined by the requirement that this property holds exactly, i.e.,
\begin{eqnarray} \label{trans1}
\widehat{A}_k^a &=& {(O_1)^a}_b A_k^b - \frac{1}{2 g} \,
{\varepsilon_{bc}}^a \left( O_1 \, \partial_k O_1^T \right)^{cb}
\nonumber \\* \widehat{\Pi}_{ka} &=& {(O_1)_a}^b \, \Pi_{kb},
\end{eqnarray}
where the orthogonal matrix $O_1$ satisfies the relation
\begin{equation} \label{gdiag1}
\widehat{G}_a = {(O_1)_a}^b G_b = \delta_{a1} \, p_1.
\end{equation}
This is clearly fulfilled if we take
\begin{subequations} \label{mat1}
\begin{eqnarray}
O_1 &=& \left( \begin{array}{ccc} \sqrt{1 - (\frac{p_2}{p_1})^2} \cos
(g \, q_2) & - \sqrt{1 - (\frac{p_2}{p_1})^2} \sin (g \, q_2) &
\frac{p_2}{p_1} \\ \frac{p_2}{p_1} \cos (g \, q_2) & - \frac{p_2}{p_1}
\sin (g \, q_2) & - \sqrt{1 - (\frac{p_2}{p_1})^2} \\ \sin (g \, q_2)
& \cos (g \, q_2) & 0 \end{array} \right) \nonumber \\* &&
\label{mat1u} \\ 
&=& \left( \begin{array}{ccc} \frac{G_1}{\sqrt{G_1^2
+ G_2^2 + G_3^2}} & \frac{G_2}{\sqrt{G_1^2 + G_2^2 + G_3^2}} &
\frac{G_3}{\sqrt{G_1^2 + G_2^2 + G_3^2}} \\[2.0ex] \frac{G_1
G_3}{\sqrt{G_1^2 + G_2^2} \sqrt{G_1^2 + G_2^2 + G_3^2}} & \frac{G_2
G_3}{\sqrt{G_1^2 + G_2^2} \sqrt{G_1^2 + G_2^2 + G_3^2}} &
-\frac{\sqrt{G_1^2 + G_2^2}}{\sqrt{G_1^2 + G_2^2 + G_3^2}} \\[2.8ex] -
\frac{G_2}{\sqrt{G_1^2 + G_2^2}} & \frac{G_1}{\sqrt{G_1^2 + G_2^2}} &
0 \end{array} \right). \nonumber \\* 
&& \label{mat1v}
\end{eqnarray}
\end{subequations}
It is interesting to note that the condition (\ref{gdiag1}) falls in
the category of Abelian gauges \cite{th}, where the gauge is partially
fixed by diagonalising some homogeneously transforming object. In our
case this object is the Gauss law generator $G = G^a \, T_a$, which is
transformed into the direction of $T_1$. The residual U(1) gauge
transformations are generated by $T_1$, and equation (\ref{gdiag1})
then suggests that $q_1$ and $p_1$ are associated with this gauge
freedom. More precisely, using the inverse formula (\ref{intrans})
together with the properties (\ref{transom}) we can calculate the
brackets
\begin{eqnarray} \label{u1rot}
\{ p_1({\bf x}), \widehat{A}_k^a({\bf y}) \} &=& - {\delta^a}_1 \,
\partial_k^{(x)} \delta({\bf x} - {\bf y}) - g \, {\varepsilon_{b1}}^a
\, \widehat{A}_k^b({\bf y}) \, \delta({\bf x} - {\bf y}), \nonumber
\\* 
\{ p_1({\bf x}), \widehat{\Pi}_{ka}({\bf y}) \} &=& - g \,
{\varepsilon^b}_{1a} \, \widehat{\Pi}_{kb}({\bf y}) \, \delta({\bf x}
- {\bf y}),
\end{eqnarray}
which prove that $p_1$ indeed generates U(1) rotations in the
direction of $T_1$. On the other hand,
$$ \{ p_1({\bf x}), \widehat{A}_k^a({\bf y}) \} = \, - \, \frac{\delta
\widehat{A}_k^a({\bf y})}{\delta q_1({\bf x})}, \qquad \{ p_1({\bf
x}), \widehat{\Pi}_{ka}({\bf y}) \} = \, - \, \frac{\delta
\widehat{\Pi}_{ka}({\bf y})}{\delta q_1({\bf x})}, $$ 
and combining these equations with the brackets (\ref{u1rot}) we get
the following functional differential equations for the fields
$\widehat{A}_k^a$ and $\widehat{\Pi}_{ka}$:
\begin{eqnarray*}
\frac{\delta \widehat{A}_k^a({\bf y})}{\delta q_1({\bf x})} &=&
{\delta^a}_1 \, \partial_k^{(x)} \delta({\bf x} - {\bf y}) + g \,
{\varepsilon_{b1}}^a \, \widehat{A}_k^b({\bf y}) \, \delta({\bf x} -
{\bf y}), \\* 
\frac{\delta \widehat{\Pi}_{ka}({\bf y})}{\delta q_1({\bf x})} &=& g
\, {\varepsilon^b}_{1a} \, \widehat{\Pi}_{kb}({\bf y}) \, \delta({\bf
x} - {\bf y}).
\end{eqnarray*}
Given that these equations hold, it is then easy to see that new
fields $\calA_k^a$ and $\calPi_{ka}$ defined by
\begin{eqnarray}
\calA_k^a &=& {(O_2)^a}_b \widehat{A}_k^b - \frac{1}{2 g} \,
{\varepsilon_{bc}}^a \left( O_2 \, \partial_k O_2^T \right)^{cb}
\nonumber \\* 
\calPi_{ka} &=& {(O_2)_a}^b \, \widehat{\Pi}_{kb},
\label{trans2} \\[1.0ex] 
O_2 &=& \left( \begin{array}{ccc} 1 & 0 & 0 \\ 0 & \cos (g \, q_1) &
-\sin (g \, q_1) \\ 0 & \sin (g \, q_1) & \cos (g \, q_1) \end{array}
\right) \label{mat2}
\end{eqnarray}
are independent of $q_1$, i.e.,
\begin{equation} \label{q1indep}
\frac{\delta \calA_k^a({\bf y})}{\delta q_1({\bf x})} = 0, \qquad
\frac{\delta \calPi_{ka}({\bf y})}{\delta q_1({\bf x})} = 0.
\end{equation}

Combining the transformations (\ref{trans1}) and (\ref{trans2}) we can
express the new variables in terms of the original fields $A_k^a$,
$\Pi_{ka}$ and the variables $\{ q_1, p_1, q_2, p_2 \}$.  Employing
formulas (\ref{transom}) and (\ref{trans}) together with the identity
$$ \{ G_a({\bf x}), q_1({\bf y}) \} = \, - \, \frac{\delta G_a({\bf
    x})}{\delta p_1({\bf y})} $$
it is then a straightforward albeit rather lengthy exercise to
check that the new variables are really gauge-invariant:
$$ \{ G_a({\bf x}), \calA_k^b({\bf y}) \} = 0, \qquad \{ G_a({\bf x}),
\calPi_{kb}({\bf y}) \} = 0. $$ 
The requirement (\ref{ginv1}) is thus satisfied, but this is not yet
sufficient to make $\calA_k^a$ and $\calPi_{ka}$ independent of
$p_1$. Moreover, the new fields are redundant in number, because they
satisfy the relation
\begin{equation} \label{gdiag2}
{\cal G}_a = \partial^{\, k} \, \calPi_{ka} - g \,
{{\varepsilon_b}^c}_a \calA^{kb} \, \calPi_{kc} = \delta_{a1} \, p_1,
\end{equation}
which is actually more like a functional identity rather than a
constraint, because it follows immediately from the transformations
(\ref{trans1}) and (\ref{trans2}). Finally, $\calA_k^a$ and
$\calPi_{ka}$ are not canonical variables due to the fact that the
gauge transformation matrices (\ref{mat1}) and (\ref{mat2}) depend on
the original fields. Employing the fundamental Poisson brackets of the
original variables and the gauge transformation properties
(\ref{transom}) it is relatively straightforward to work out the
brackets of $\calA_k^a$ and $\calPi_{ka}$, but the calculations are
lengthy. In fact, it becomes almost inevitable to use computer
software capable of symbolic manipulations to perform these extensive
calculations.  Eventually we obtain the following result:
\begin{subequations} \label{pbr1}
\begin{eqnarray}
\{ \calA_k^a({\bf x}), \calPi_{lb}({ \bf y}) \} &=& \frac{1}{p_1({ \bf
y})} \Bigl[ {\delta^a}_b \, \calPi_{l1}({ \bf y}) - \delta_{b1} \,
\calPi_l^a({ \bf y}) \Bigr] \, \partial_k^{(x)} \delta({\bf x} - {\bf
y}) \nonumber \\*[1.0ex] 
&& + \Bigl[ \frac{g}{2 \, p_1({ \bf y})} \Bigl( {\delta^a}_1 \,
{\varepsilon_{bc}}^d + \delta_{b1} \, {{\varepsilon^a}_c}^d +
{\delta^d}_1 \, {\varepsilon^a}_{bc} + \delta_{c1} \,
{{\varepsilon^a}_b}^d \Bigr) \nonumber \\* 
&& \hspace{4.4em} \times \calA_k^c({ \bf y}) \calPi_{ld}({ \bf y}) +
\delta_{kl} \, {\delta^a}_b \Bigr] \, \delta({\bf x} - {\bf y})
\nonumber \\[1.0ex] 
&& + \left[ {\delta^a}_1 \, \partial_k^{(x)} - g \,
{\varepsilon_{c1}}^a \calA_k^c({\bf x}) \right] \frac{\delta
\calPi_{lb}({\bf y})}{\delta p_1({\bf x})} \nonumber \\*[1.0ex] 
&& + g \, {\varepsilon^c}_{1b} \, \calPi_{lc}({\bf y}) \frac{\delta
\calA_k^a({\bf x})}{\delta p_1({\bf y})}, \label{pbr1ap} \\[1.0ex] 
\{ \calA_k^a({\bf x}), \calA_l^b({ \bf y}) \} &=& \frac{1}{g} \,
{\varepsilon_1}^{ab} \, \partial_l^{(y)} \left[ \frac{1}{p_1({ \bf
y})} \, \partial_k^{(x)} \delta({\bf x} - {\bf y}) \right] \nonumber
\\*[1.0ex] 
&& + \frac{1}{p_1({ \bf y})} \Bigl[ \delta^{ab} \, \calA_{l1}({ \bf
y}) - {\delta^b}_1 \, \calA_l^a ({ \bf y}) \Bigr] \, \partial_k^{(x)}
\delta({\bf x} - {\bf y}) \nonumber \\[1.0ex] 
&& - \partial_l^{(y)} \left[ \frac{1}{p_1({ \bf y})} \Bigl(
\delta^{ab} \, \calA_{k1}({ \bf y}) - {\delta^a}_1 \, \calA_k^b ({ \bf
y}) \Bigr) \, \delta({\bf x} - {\bf y}) \right] \nonumber \\[1.0ex] 
&& + \frac{g}{2 \, p_1({ \bf y})} \Bigl( {\delta^a}_1 \,
{\varepsilon^b}_{cd} + {\delta^b}_1 \, {\varepsilon^a}_{cd} +
\delta_{d1} \, {\varepsilon^{ab}}_c + \delta_{c1} \,
{\varepsilon^{ab}}_d \Bigr) \nonumber \\* 
&& \hspace{4em} \times \calA_k^c({ \bf y}) \calA_l^d({ \bf y}) \,
\delta({\bf x} - {\bf y}) \nonumber \\[1.0ex] 
&& + \left[ {\delta^a}_1 \, \partial_k^{(x)} - g \,
{\varepsilon_{c1}}^a \calA_k^c({\bf x}) \right] \frac{\delta
\calA_l^b({\bf y})}{\delta p_1({\bf x})} \nonumber \\*[1.0ex] 
&& - \left[ {\delta^b}_1 \, \partial_l^{(y)} - g \,
{\varepsilon_{c1}}^b \calA_l^c({\bf y}) \right] \frac{\delta
\calA_k^a({\bf x})}{\delta p_1({\bf y})},
\label{pbr1aa}  \\[1.0ex]
\{ \calPi_{ka}({\bf x}), \calPi_{lb}({ \bf y}) \} &=& \frac{g}{2 \,
p_1({ \bf y})} \Bigl( \delta_{a1} \, {\varepsilon_b}^{cd} +
\delta_{b1} \, {\varepsilon_a}^{cd} + {\delta^d}_1 \,
{\varepsilon_{ab}}^c + {\delta^c}_1 \, {\varepsilon_{ab}}^d \Bigr)
\hspace{3.1em} \nonumber \\*[0.5ex] 
&& \hspace{0.2em} \times \calPi_{kc}({ \bf y}) \calPi_{ld}({ \bf y})
\, \delta({\bf x} - {\bf y}) - g \, {\varepsilon^c}_{1a} \,
\calPi_{kc}({\bf x}) \frac{\delta \calPi_{lb}({\bf y})}{\delta
p_1({\bf x})} \nonumber \\*[0.5ex] 
&& + g \, {\varepsilon^c}_{1b} \, \calPi_{lc}({\bf y}) \frac{\delta
\calPi_{ka}({\bf x})}{\delta p_1({\bf y})}.
\label{pbr1pp}
\end{eqnarray}
\end{subequations}
As there are redundant coordinates in this set of variables, we should
verify that these brackets are compatible with equation
(\ref{gdiag2}). Indeed, starting from the brackets (\ref{pbr1}) it is
possible to derive the result
$$ \{ {\cal G}_a({\bf x}), \calA_k^b({\bf y}) \} = 0, \qquad \{ {\cal
G}_a({\bf x}), \calPi_{kb}({\bf y}) \} = 0, $$ 
which is consistent with equations (\ref{q1indep}) and
(\ref{gdiag2}). Our next task is to pa\-ra\-met\-rise $\calA_k^a$ and
$\calPi_{ka}$ with new canonical variables in such a way that the
relations (\ref{gdiag2}) and (\ref{pbr1}) are satisfied.

\subsection{Canonical variables} \label{canvar}

The elimination of the residual U(1) gauge degree of freedom with the
transformation (\ref{trans2}) was rather symbolic by nature, because
the form of $q_1$ was not specified. The advantage of doing so is the
fact that the Poisson brackets (\ref{pbr1}) now hold for all possible
U(1) gauges and we can thus experiment with different gauge
choices. Once a choice is made, its consistency with the brackets
(\ref{pbr1}) then yields equations that determine the $p_1$-dependence
of $\calA_k^a$ and $\calPi_{ka}$. The ingredients for choosing the
gauge can be read off from the transformation formula
(\ref{trans2}). It is seen that the available objects fall in three
categories: the pairs $(\widehat{A}_k^2, \widehat{A}_k^3)$ and
$(\widehat{\Pi}_{k2}, \widehat{\Pi}_{k3})$ form SO(2) doublets, the
components $\widehat{\Pi}_{k1}$ are invariant and $\widehat{A}_k^1$
transforms as a photon. Although every gauge is possible from the
physical point of view, yet in practice some gauges are not manifestly
compatible with the brackets (\ref{pbr1}). For example, in the Coulomb
gauge we should choose $q_1$ in such a way that the equation
$\partial^k \calA_k^1 = 0$ would hold as a functional identity, as
indicated by equations (\ref{q1indep}). Therefore the Poisson brackets
$\{ \partial^k \calA_k^1({\bf x}), \partial^l \calA_l^1({ \bf y}) \}$
should also vanish identically, but according to the relations
(\ref{pbr1aa}) this is not the case. I cannot give a definite reason
for this contradiction, but the canonical structure of the variables
already fixed might be the cause and the problem could possibly be
circumvented by adjusting $q_1$ and the definitions (\ref{intrans})
suitably.

In the following calculations I have chosen the unitary gauge
$$ \calPi_{12}({ \bf x}) = 0. $$
Its consistency with the resulting identity $\{ \calPi_{12}({\bf x}),
\calPi_{12}({ \bf y}) \} = 0$ is obvious and it corresponds to
defining $q_1$ by
\begin{equation} \label{q1def}
q_1 = \frac{2}{g} \arctan \left( \frac{\sqrt{\widehat{\Pi}_{12}^2 +
\widehat{\Pi}_{13}^2} - \widehat{\Pi}_{13}}{\widehat{\Pi}_{12}}
\right).
\end{equation}
Using formulas (\ref{transom}), (\ref{intrans}) and (\ref{mat1v}) it
is a straightforward but lengthy calculation to verify that the
fundamental Poisson bracket relations between $q_1$ and the variables
$\{p_1,q_2,p_2\}$ indeed hold. Now the functional identities
$$ \{ \calA_k^a({\bf x}), \calPi_{12}({ \bf y}) \} = 0, \qquad \{
\calPi_{ka}({\bf x}), \calPi_{12}({ \bf y}) \} = 0 $$ 
combined with the brackets (\ref{pbr1}) give the following equations:
\begin{eqnarray*}
\frac{\delta \calA_k^a({\bf x})}{\delta p_1({\bf y})} &=& - \,
\frac{1}{g \, p_1({ \bf y})} \frac{\calPi_{11}({\bf
y})}{\calPi_{13}({\bf y})} \, {\delta^a}_2 \, \partial_k^{(x)}
\delta({\bf x} - {\bf y}) \\*[1.0ex] 
&& + \, \frac{1}{\calPi_{13}({\bf y})} \left( \frac{\calPi_{11}({\bf
y})}{p_1({\bf y})} \, {\varepsilon_{b2}}^a \calA_k^b({\bf y}) -
\frac{1}{g} \, \delta_{k1} \, {\delta^a}_2 \right) \delta({\bf x} -
{\bf y}), \\[1.0ex] 
\frac{\delta \calPi_{ka}({\bf x})}{\delta p_1({\bf y})} &=&
\frac{1}{p_1({ \bf y})} \frac{\calPi_{11}({\bf y})}{\calPi_{13}({\bf
y})} \, {\varepsilon^b}_{2a} \, \calPi_{kb}({\bf y}) \, \delta({\bf x}
- {\bf y}).
\end{eqnarray*}
At first sight these equations look a bit frightening, but they turn
out to be solvable with a reasonable effort. The solution can be
written as a gauge transformation in the direction of $T_2$:
\begin{eqnarray} \label{intrans3}
\calA_k^a &=& {(O_3)^a}_b \left( {\cal Q}_k^b - \frac{1}{g} \,
\delta_{k1} \, {\delta^b}_2 \frac{p_1}{{\cal P}_{11}} \cos \phi
\right) - \frac{1}{2 g} \, {\varepsilon_{bc}}^a \left( O_3 \,
\partial_k O_3^T \right)^{cb} \nonumber \\* 
\calPi_{ka} &=& {(O_3)_a}^b \, {\cal P}_{kb}
\end{eqnarray}
with
\begin{eqnarray}
O_3 &=& \left( \begin{array}{ccc} \sin \phi & 0 & - \cos \phi \\ 0 &
 -1 & 0 \\ -\cos \phi & 0 & - \sin \phi \end{array} \right), \nonumber
 \\[1.0ex] 
&& \sin \phi = \frac{\beta}{p_1}, \label{phidef}
\end{eqnarray}
where ${\cal Q}_k^a$, ${\cal P}_{ka}$ and $\beta$ are constants of
integration, i.e.,
$$ \frac{\delta {\cal Q}_k^a({\bf x})}{\delta p_1({\bf y})} = 0,
\qquad \frac{\delta {\cal P}_{ka}({\bf x})}{\delta p_1({\bf y})} = 0,
\qquad \frac{\delta \beta({\bf x})}{\delta p_1({\bf y})} = 0, $$ 
and
\begin{equation} \label{p1zero}
{\cal P}_{12}({\bf x}) = {\cal P}_{13}({\bf x}) = 0.
\end{equation}
These constant fields fulfill both of the requirements (\ref{ginv}),
and therefore they should be adopted as new variables. The
transformation formula is the inverse of equation (\ref{intrans3}),
that is,
\begin{eqnarray} \label{trans3}
{\cal Q}_k^a &=& {(O_3)^a}_b \left( \calA_k^b + \frac{1}{g} \,
\delta_{k1} \, {\delta^b}_2 \, p_1 \, \frac{\calPi_{13}}{\calPi_{11}^2
+ \calPi_{13}^2} \right) - \frac{1}{2 g} \, {\varepsilon_{bc}}^a
\left( O_3 \, \partial_k O_3^T \right)^{cb} \nonumber \\* 
{\cal P}_{ka} &=& {(O_3)_a}^b \, \calPi_{kb},
\end{eqnarray}
where we now write the gauge angle as
\begin{equation} \label{phidef+}
\sin \phi = \frac{\calPi_{11}}{\sqrt{\calPi_{11}^2 + \calPi_{13}^2}},
\qquad \cos \phi = \, - \, \frac{\calPi_{13}}{\sqrt{\calPi_{11}^2 +
\calPi_{13}^2}}.
\end{equation}
Note that the matrix $O_3$ is both orthogonal and symmetric.

In order to proceed towards our final canonical transformation we must
now find out whether the newest set of variables can be made canonical
in accordance with the relation (\ref{gdiag2}). Using equations
(\ref{intrans3}) it is easy to see that the corresponding relation in
the new variables reads
\begin{equation} \label{gdiag3}
\partial^{\, k} \, {\cal P}_{ka} - g \, {{\varepsilon_b}^c}_a {\cal
  Q}^{kb} \, {\cal P}_{kc} = \delta_{a1} \, \beta.
\end{equation}
The Poisson brackets are evaluated by inserting expressions
(\ref{trans3}) into the relations (\ref{pbr1}). Again this is a
formidable calculation which requires extensive use of computer
software. Here is the result:
\begin{eqnarray} \label{pbr2}
\{ {\cal Q}_k^a({\bf x}), {\cal P}_{lb}({ \bf y}) \} &=& \delta_{kl}
\, {\delta^a}_b \, \delta({\bf x} - {\bf y}), \quad k \neq 1 \nonumber
\\*[1.8ex] 
\{ {\cal Q}_1^a({\bf x}), {\cal P}_{11}({ \bf y}) \} &=& {\delta^a}_1
\, \delta({\bf x} - {\bf y}) \nonumber \\[1.2ex] 
\{ {\cal Q}_1^a({\bf x}), {\cal P}_{lb}({ \bf y}) \} &=& - \,
\frac{1}{{\cal P}_{11}({ \bf y})} \left[ {\delta^a}_b {\cal P}_{l1}({
\bf y}) - \delta_{b1} {\cal P}_l^a({ \bf y}) \right] \delta({\bf x} -
{\bf y}), \quad l \neq 1 \nonumber \\[0.7ex] 
\{ {\cal Q}_k^a({\bf x}), {\cal Q}_l^b({ \bf y}) \} &=& 0, \quad k
\neq 1, \hspace{0.8em} l \neq 1 \nonumber \\[1.0ex]
\{ {\cal Q}_k^a({\bf x}), {\cal Q}_1^b({ \bf y}) \} &=& - \,
\frac{1}{g} \, {\varepsilon_1}^{ab} \frac{1}{{\cal P}_{11}({ \bf y})}
\, \partial_k^{(x)} \delta({\bf x} - {\bf y}) \\* 
&& + \, \frac{1}{{\cal P}_{11}({ \bf y})} \left[ \delta^{ab} {\cal
Q}_{k1}({ \bf y}) - {\delta^a}_1 {\cal Q}_k^b({ \bf y}) \right]
\delta({\bf x} - {\bf y}), \quad k \neq 1 \nonumber \\[0.5ex] 
\{ {\cal Q}_1^a({\bf x}), {\cal Q}_1^b({ \bf y}) \} &=& - \,
\frac{1}{{\cal P}_{11}({ \bf y})} \Bigl[ {\delta^a}_1 \, {\cal
Q}_1^b({ \bf y}) - {\delta^b}_1 \, {\cal Q}_1^a({ \bf y}) \nonumber
\\* 
&& \hspace{4.5em} + \, \frac{1}{g} \, {\varepsilon_1}^{ab} \,
\frac{\beta({ \bf y}) - \partial_1^{(y)} {\cal P}_{11}({ \bf
y})}{{\cal P}_{11}({ \bf y})} \Bigr] \, \delta({\bf x} - {\bf y})
\nonumber \\ 
\{ {\cal P}_{ka}({\bf x}), {\cal P}_{lb}({ \bf y}) \} &=& 0 \nonumber
\\[1.5ex] 
\{ \beta({\bf x}), {\cal Q}_k^a({ \bf y}) \} &=& - {\delta^a}_1 \,
\partial_k^{(x)} \delta({\bf x} - {\bf y}) - g \, {\varepsilon_{b1}}^a
\, {\cal Q}_k^b({\bf y}) \, \delta({\bf x} - {\bf y}) \nonumber
\\*[1.6ex] 
\{ \beta({\bf x}), {\cal P}_{ka}({ \bf y}) \} &=& - g \, (1 -
\delta_{k1}) \, {\varepsilon^b}_{1a} \, {\cal P}_{kb}({ \bf y}) \,
\delta({\bf x} - {\bf y}). \nonumber
\end{eqnarray}
Remember that ${\cal P}_{12} = {\cal P}_{13} = 0$ in these
relations. It is now easy to construct the desired canonical
fields. The brackets (\ref{pbr2}) suggest that we choose the pairs
\begin{equation} \label{canpairs}
({\cal Q}_1^1, {\cal P}_{11}), \qquad ({\cal Q}_k^a, {\cal P}_{ka}),
\quad k \neq 1
\end{equation}
as canonically conjugate variables. If we then solve the remaining
variables from equation (\ref{gdiag3}),
\begin{subequations} \label{redu1}
\begin{eqnarray}
\beta &=& \partial_1 \, {\cal P}_{11} + \sum_{k=2}^3 \left( \partial_k
\, {\cal P}_{k1} - g \, {{\varepsilon_b}^c}_1 {\cal Q}_k^b \, {\cal
P}_{kc} \right) \label{redu1a} \\ 
{\cal Q}_1^2 &=& - \, \frac{1}{g \, {\cal P}_{11}} \sum_{k=2}^3 \left(
\partial_k \, {\cal P}_{k3} - g \, {{\varepsilon_b}^c}_3 {\cal Q}_k^b
\, {\cal P}_{kc} \right) \\ 
{\cal Q}_1^3 &=& \frac{1}{g \, {\cal P}_{11}} \sum_{k=2}^3 \left(
\partial_k \, {\cal P}_{k2} - g \, {{\varepsilon_b}^c}_2 {\cal Q}_k^b
\, {\cal P}_{kc} \right),
\end{eqnarray}
\end{subequations}
it turns out that all the Poisson brackets in the set (\ref{pbr2})
involving these variables follow from the fundamental brackets of the
pairs (\ref{canpairs}). Unfortunately the variables (\ref{canpairs}),
although gauge-invariant and canonical, are still not useful for
implementing the Gauss law. The reason is equation (\ref{phidef}),
which shows that $\beta$ tends to zero in the limit when $p_1$
vanishes. Looking at expression (\ref{redu1a}), we see that it would
be difficult to implement the requirement $\beta \rightarrow 0$ using
the variables (\ref{canpairs}). A suitable canonical transformation is
needed.

\subsection{Canonical U(1) transformation}

Passing to the variables (\ref{canpairs}), we have replaced the
original SU(2) fields with a set of gauge-invariant canonical
fields. However, the Poisson brackets (\ref{pbr2}) show that these
variables have an inner U(1) symmetry, which is generated by
$\beta$. Note that this symmetry has nothing to do with the original
SU(2) symmetry since all the variables (\ref{canpairs}) and the
generator $\beta$, defined by equation (\ref{redu1a}), are
gauge-invariant with respect to the generators $G_a$. Even so, we can
apply the procedures of sections \ref{gpar} -- \ref{canvar} also to
this U(1) symmetry and choose $\beta$ as a new canonical momentum
variable, i.e.,
\begin{equation} \label{p3def}
p_3 = \partial_1 \, {\cal P}_{11} + \sum_{k=2}^3 \left( \partial_k \,
{\cal P}_{k1} - g \, {{\varepsilon_b}^c}_1 {\cal Q}_k^b \, {\cal
P}_{kc} \right).
\end{equation}
The canonical conjugate of $p_3$ then determines the gauge angle
associated with transformations generated by $p_3$, but again we leave
the specific form of $q_3$ open at this stage. Since both $q_3$ and
$p_3$ must have vanishing Poisson brackets with the remaining
variables of the final canonical set, we conclude that the remaining
variables must be functionally independent of $q_3$ and $p_3$. The
elimination of $q_3$ can be done, as before, with a gauge
transformation in the $T_1$-direction:
\begin{eqnarray}
\widehat{Q}_k^a &=& {(O_4)^a}_b {\cal Q}_k^b - \frac{1}{2 g} \,
{\varepsilon_{bc}}^a \left( O_4 \, \partial_k O_4^T \right)^{cb}
\nonumber \\* 
\widehat{P}_{ka} &=& {(O_4)_a}^b \, {\cal P}_{kb},
\label{trans4}
\end{eqnarray}
where
$$ O_4 = \left( \begin{array}{ccc} 1 & 0 & 0 \\ 0 & \cos (g \, q_3) &
-\sin (g \, q_3) \\ 0 & \sin (g \, q_3) & \cos (g \, q_3) \end{array}
\right). $$ 
The Poisson brackets of the new variables follow from the algebra
(\ref{pbr2}), the result being
\begin{eqnarray} \label{pbr3}
\{ \widehat{Q}_k^a({\bf x}), \widehat{P}_{lb}({ \bf y}) \} &=&
\delta_{kl} \, {\delta^a}_b \, \delta({\bf x} - {\bf y}) + \left[
{\delta^a}_1 \, \partial_k^{(x)} - g \, {\varepsilon_{c1}}^a
\widehat{Q}_k^c({\bf x}) \right] \frac{\delta \widehat{P}_{lb}({\bf
y})}{\delta p_3({\bf x})} \nonumber \\* && 
+ \, g \, {\varepsilon^c}_{1b} \, \widehat{P}_{lc}({\bf y})
\frac{\delta \widehat{Q}_k^a({\bf x})}{\delta p_3({\bf y})}, \nonumber
\\ 
\{ \widehat{Q}_k^a({\bf x}), \widehat{Q}_l^b({ \bf y}) \} &=& \left[
{\delta^a}_1 \, \partial_k^{(x)} - g \, {\varepsilon_{c1}}^a
\widehat{Q}_k^c({\bf x}) \right] \frac{\delta \widehat{Q}_l^b({\bf
y})}{\delta p_3({\bf x})} \\*[0.5ex] 
&& - \left[ {\delta^b}_1 \, \partial_l^{(y)} - g \,
{\varepsilon_{c1}}^b \widehat{Q}_l^c({\bf y}) \right] \frac{\delta
\widehat{Q}_k^a({\bf x})}{\delta p_3({\bf y})}, \nonumber \\* 
\{ \widehat{P}_{ka}({\bf x}), \widehat{P}_{lb}({ \bf y}) \} &=& - g \,
{\varepsilon^c}_{1a} \, \widehat{P}_{kc}({\bf x}) \frac{\delta
\widehat{P}_{lb}({\bf y})}{\delta p_3({\bf x})} + g \,
{\varepsilon^c}_{1b} \, \widehat{P}_{lc}({\bf y}) \frac{\delta
\widehat{P}_{ka}({\bf x})}{\delta p_3({\bf y})}.  \nonumber
\end{eqnarray}
For the sake of clarity I have written down only those brackets that
hold for the actual variables
$$ (\widehat{Q}_1^1, \widehat{P}_{11}), \qquad (\widehat{Q}_k^a,
\widehat{P}_{ka}), \quad k \neq 1. $$

In order to define variables independent of $p_3$ we must now specify
the U(1) gauge by fixing $q_3$. I have chosen
\begin{equation} \label{q3def}
q_3 = \frac{2}{g} \arctan \left( \frac{\sqrt{{\cal P}_{22}^2 + {\cal
P}_{23}^2} - {\cal P}_{23}}{{\cal P}_{22}} \right),
\end{equation}
which corresponds to the identity
\begin{equation} \label{p2zero}
\widehat{P}_{22}({\bf x}) = 0.
\end{equation}
Making use of the brackets (\ref{pbr2}) it is possible to verify that
$q_3$ and $p_3$ indeed satisfy
$$ \{ q_3({\bf x}), p_3({\bf y}) \} = \delta({\bf x} - {\bf y}), $$
while their brackets with the variables $\{q_1,p_1,q_2,p_2\}$ vanish
due to the fact that both ${\cal Q}_k^a$ and ${\cal P}_{ka}$ meet the
requirements (\ref{ginv}). As before, the functional identities
$$ \{ \widehat{Q}_k^a({\bf x}), \widehat{P}_{22}({ \bf y}) \} = 0, \qquad
\{ \widehat{P}_{ka}({\bf x}), \widehat{P}_{22}({ \bf y}) \} = 0 $$ 
lead to the equations
\begin{eqnarray*}
\frac{\delta \widehat{Q}_k^a({\bf x})}{\delta p_3({\bf y})} &=& - \,
\frac{1}{g \, \widehat{P}_{23}({ \bf y})} \, \delta_{k2} \,
{\delta^a}_2 \, \delta({\bf x} - {\bf y}), \\* 
\frac{\delta \widehat{P}_{ka}({\bf x})}{\delta p_3({\bf y})} &=& 0,
\end{eqnarray*}
whose solutions read
\begin{eqnarray} \label{intrans5}
\widehat{Q}_k^a &=& Q_k^a \, - \, \frac{1}{g} \, \delta_{k2} \,
{\delta^a}_2 \, \frac{p_3}{P_{23}} \nonumber \\* 
\widehat{P}_{ka} &=& P_{ka},
\end{eqnarray}
where $Q_k^a$ and $P_{ka}$ are constants of integration, i.e.,
$$ \frac{\delta Q_k^a({\bf x})}{\delta p_3({\bf y})} = 0, \qquad
\frac{\delta P_{ka}({\bf x})}{\delta p_3({\bf y})} = 0. $$ 
Now we choose these constants as new variables. Equation (\ref{p3def})
then leads to the relation
\begin{equation} \label{u1gs}
\partial_1 \, P_{11} + \sum_{k=2}^3 \left( \partial_k \, P_{k1} - g \,
{{\varepsilon_b}^c}_1 Q_k^b \, P_{kc} \right) = 0,
\end{equation}
which holds as a functional identity, implying that the new variables
contain one redundant coordinate. The Poisson brackets of $Q_k^a$ and
$P_{ka}$ are easily evaluated with the help of the relations
(\ref{pbr3}). The result reads
\begin{eqnarray} \label{pbr4}
\{ Q_k^a({\bf x}), P_{lb}({ \bf y}) \} &=& \delta_{kl} \, {\delta^a}_b
\, \delta({\bf x} - {\bf y}) - \delta_{k2} \, {\delta^a}_2 \,
{\varepsilon^c}_{1b} \, \frac{1}{P_{23}({\bf y})} P_{lc}({\bf y}) \,
\delta({\bf x} - {\bf y}) \nonumber \\*[0.5ex] 
\{ Q_k^a({\bf x}), Q_l^b({ \bf y}) \} &=& 0, \qquad (k,a) \neq (2,2),
\hspace{0.8em} (l,b) \neq (2,2) \nonumber \\[1.0ex]
\{ Q_k^a({\bf x}), Q_2^2({ \bf y}) \} &=& - \, \frac{1}{g \, P_{23}({
\bf y})} \left[ {\delta^a}_1 \, \partial_k^{(x)} - g \,
{\varepsilon_{c1}}^a \, Q_k^c({\bf x}) \right] \delta({\bf x} - {\bf
y}), \nonumber \\*[0.8ex] && \hspace{3em} (k,a) \neq (2,2) \\* 
\{ Q_2^2({\bf x}), Q_2^2({ \bf y}) \} &=& 0 \nonumber \\*[1.7ex] 
\{ P_{ka}({\bf x}), P_{lb}({ \bf y}) \} &=& 0, \nonumber
\end{eqnarray}
showing that the pairs
$$ (Q_1^1, P_{11}), \qquad (Q_2^a, P_{2a}), \hspace{0.8em} a = 1,3,
\qquad (Q_3^a, P_{3a}), \hspace{0.8em}a = 1,2,3 $$ 
are the most natural choice for final canonical variables. Solving
equation (\ref{u1gs}) for the redundant coordinate,
\begin{equation} \label{Q22def}
Q_2^2 = \frac{1}{g \, P_{23}} \left( \partial^{\, k} \, P_{k1} - g \,
{{\varepsilon_b}^c}_1 \, Q_3^b \, P_{3c} \right),
\end{equation}
it is easy to see that all of the Poisson bracket relations
(\ref{pbr4}) hold true. Our search for suitable canonical variables is
now over.

\subsection{Results}

Starting from the original canonical fields $(A_k^a, \Pi_{ka})$ and
passing through four sets of intermediate variables we have found the
final canonical pairs
\begin{equation} \label{finvar}
\begin{array}{ll}
(q_i, p_i), & i = 1,2,3 \\[0.5ex] 
(Q_1^1, P_{11}) & \\[0.5ex] 
(Q_2^a, P_{2a}), & a = 1,3 \\[0.5ex] 
(Q_3^a, P_{3a}), & a = 1,2,3.
\end{array}
\end{equation}
Equation (\ref{intrans}) relates $p_1$, $q_2$ and $p_2$ to the
original variables, and a formula for $q_1$ is obtained by combining
equations (\ref{q1def}), (\ref{trans1}) and (\ref{mat1v}). The
momentum $p_3$ is most easily calculated by combining equations
(\ref{phidef}), (\ref{phidef+}), (\ref{trans2}) and (\ref{trans1}),
while it takes successive applications of equations (\ref{q3def}),
(\ref{trans3}), (\ref{trans2}) and (\ref{trans1}) to work out a
formula for $q_3$.  The remaining variables of the set (\ref{finvar})
are then obtained by performing the transformations (\ref{intrans5}),
(\ref{trans4}), (\ref{trans3}), (\ref{trans2}) and (\ref{trans1}) one
after the other. Again the manipulations are so lengthy that computer
assistance is required. Introducing the notation
$$ ||X|| := \sqrt{X_a X^a} $$ 
for the Lie algebra norm, the results can be written as follows:
\begin{subequations} \label{fintrans}
\begin{eqnarray}
q_1 &=& \frac{2}{g} \arctan \left( \frac{\sqrt{\widehat{\Pi}_{12}^2 +
\widehat{\Pi}_{13}^2} - \widehat{\Pi}_{13}}{\widehat{\Pi}_{12}}
\right), \label{ftq1} \\*[1.0ex] 
&& \hspace{1em} \widehat{\Pi}_{12} = \frac{1}{||G||} \Biggl[
\frac{G_3}{\sqrt{G_1^2 + G_2^2}} \, \left( G_1 \, \Pi_{11} + G_2 \,
\Pi_{12} \right) - \sqrt{G_1^2 + G_2^2} \, \, \Pi_{13} \Biggr]
\nonumber \\ 
&& \hspace{1em} \widehat{\Pi}_{13} = \frac{1}{\sqrt{G_1^2 + G_2^2}} \,
\left( -G_2 \, \Pi_{11} + G_1 \, \Pi_{12} \right) \nonumber \\[1.5ex]
q_2 &=& - \, \frac{2}{g} \arctan \left( \frac{\sqrt{G_1^2 + G_2^2} -
G_1}{G_2} \right) \label{ftq2} \\[1.5ex] 
q_3 &=& \frac{2}{g} \arctan \left( \frac{\sqrt{{\cal P}_{22}^2 + {\cal
P}_{23}^2} - {\cal P}_{23}}{{\cal P}_{22}} \right),
\label{ftq3} \\[1.5ex] 
&& \hspace{1em} {\cal P}_{22} = \frac{1}{||{\cal N}||} \,
\varepsilon^{abc} \, G_a \Pi_{1b} \Pi_{2c} \nonumber \\[1.5ex] 
&& \hspace{1em} {\cal P}_{23} = \frac{1}{||{\cal N}||} \, \frac{1}{||
\Pi_1 ||} \left( \delta^{ad} \, \delta^{bc} - \delta^{ab} \,
\delta^{cd} \right) G_a \Pi_{1b} \Pi_{1c} \Pi_{2d} \nonumber \\[1.5ex]
&& \hspace{1em} {\cal N}_a = {\varepsilon_a}^{bc} \, G_b \, \Pi_{1c}
\nonumber \\[1.5ex] 
p_1 &=& ||G|| \label{ftp1} \\[1.8ex] 
p_2 &=& G_3 \label{ftp2} \\[1.8ex] 
p_3 &=& \frac{G^a \, \Pi_{1a}}{|| \Pi_1 ||} \label{ftp3} \\[1.0ex] 
Q_k^a &=& {\Omega^a}_b A_k^b - \frac{1}{2 g} \, {\varepsilon_{bc}}^a
\left( \Omega \, \partial_k \Omega^T \right)^{cb} \label{ftQ} \\* 
P_{ka} &=& {\Omega_a}^b \, \Pi_{kb}, \label{ftP}
\end{eqnarray}
\end{subequations}
where
\begin{eqnarray} \label{tep}
{\Omega_a}^b &=& {\left( O_4 O_3 O_2 O_1 \right)_a}^b \nonumber
\\[1.0ex] 
&=& \delta_{a1} \, \frac{1}{|| \Pi_1 ||} \, \Pi_1^b + \delta_{a2} \,
\frac{1}{||N||} \, \varepsilon^{bcd} \, \Pi_{2c} \Pi_{1d} \nonumber
\\*[1.0ex] 
&& + \, \delta_{a3} \, \frac{1}{|| \Pi_1 ||} \, \frac{1}{||N||} \left(
\delta^{be} \, \delta^{cd} - \delta^{bc} \, \delta^{de} \right)
\Pi_{1c} \Pi_{1d} \Pi_{2e}, \\*[1.0ex] 
N_a &=& {\varepsilon_a}^{bc} \, \Pi_{1b} \Pi_{2c}. \nonumber
\end{eqnarray}
This transformation is singular when $|| \Pi_1 ||$ or $||N||$
vanishes, corresponding to points where the gauge angles
(\ref{phidef+}) and (\ref{ftq3}) become ambiguous. These singularities
are Gribov ambiguities \cite{g} peculiar to unitary gauges, and it is
well known that such ambiguities appear in almost every gauge
\cite{s}.

When inverting the transformation (\ref{fintrans}) it should be noted
that formula (\ref{ftQ}) holds for variables of the set (\ref{finvar})
only. The general expression reads
$$ Q_k^a + \frac{1}{g} \, \delta_{k1} \, {(O_4)^a}_2
\frac{p_1}{P_{11}} \sqrt{1 - \left( \frac{p_3}{p_1} \right)^2} -
\frac{1}{g} \, \delta_{k2} \, {\delta^a}_2 \frac{p_3}{P_{23}} \, = \,
{\Omega^a}_b A_k^b - \frac{1}{2 g} \, {\varepsilon_{bc}}^a \left(
\Omega \, \partial_k \Omega^T \right)^{cb}. $$ 
This equation determines the original gauge potential $A_k^a$ as a
function of the variables (\ref{finvar}), provided that we use
equations (\ref{redu1}), (\ref{Q22def}), (\ref{trans4}) and
(\ref{intrans5}) to define those components $Q_k^a$ that are not
regarded as free variables. In the same way we can invert the momentum
transformation equation (\ref{ftP}), taking into account the
definitions (\ref{p1zero}) and (\ref{p2zero}). The result is
\begin{subequations} \label{finintr}
\begin{eqnarray}
A_k^a &=& {\left( \Omega^T \right)^a}_b \left( Q_k^b + \frac{1}{g} \,
\delta_{k1} \, {(O_4)^b}_2 \frac{p_1}{P_{11}} \sqrt{1 - \left(
\frac{p_3}{p_1} \right)^2} \, - \, \frac{1}{g} \, \delta_{k2} \,
{\delta^b}_2 \frac{p_3}{P_{23}} \right)\nonumber \\* && 
- \frac{1}{2 g} \, {\varepsilon_{bc}}^a \left( \Omega^T \, \partial_k
\Omega \right)^{cb} \label{finintra} \\* 
\Pi_{ka} &=& {\left( \Omega^T \right)_a}^b \, P_{kb}, \label{finintrp}
\end{eqnarray}
\end{subequations}
where
\begin{eqnarray} \label{konsta}
Q_1^2 &=& - \, \frac{1}{g \, P_{11}} \sum_{k=2}^3 \left( \partial_k \,
P_{k3} - g \, {{\varepsilon_b}^c}_3 Q_k^b \, P_{kc} \right) +
\frac{1}{g} \, p_3 \, \frac{P_{21}}{P_{11} P_{23}} \nonumber \\* 
Q_1^3 &=& \frac{1}{g \, P_{11}} \Bigl( \partial_3 \, P_{32} -
\sum_{k=2}^3 g \, {{\varepsilon_b}^c}_2 Q_k^b \, P_{kc} \Bigr) \\
Q_2^2 &=& \frac{1}{g \, P_{23}} \left( \partial^{\, k} \, P_{k1} - g
\, {{\varepsilon_b}^c}_1 \, Q_3^b \, P_{3c} \right) \nonumber
\\*[2.0ex] 
P_{12} &=& P_{13} \hspace{0.3em} = \hspace{0.3em} P_{22}
\hspace{0.3em} = \hspace{0.3em} 0 \nonumber
\end{eqnarray}
and $\Omega^T$ is expressed in the variables $(q_i, p_i)$, i.e.,
\begin{eqnarray*}
\Omega^T &=& \left( \begin{array}{ccc} \sqrt{1 - (\frac{p_2}{p_1})^2}
\cos (g \, q_2) & \frac{p_2}{p_1} \cos (g \, q_2) & \sin (g \, q_2) \\
- \sqrt{1 - (\frac{p_2}{p_1})^2} \sin (g \, q_2) & - \frac{p_2}{p_1}
\sin (g \, q_2) & \cos (g \, q_2) \\ \frac{p_2}{p_1} & - \sqrt{1 -
(\frac{p_2}{p_1})^2} & 0 \end{array} \right) \\* 
&& \cdot \left( \begin{array}{ccc} 1 & 0 & 0 \\ 0 & \cos (g \, q_1) &
\sin (g \, q_1) \\ 0 & -\sin (g \, q_1) & \cos (g \, q_1) \end{array}
\right) \\ 
&& \cdot \left( \begin{array}{ccc} \frac{p_3}{p_1} & 0 & \sqrt{1 -
(\frac{p_3}{p_1})^2} \\ 0 & -1 & 0 \\ \sqrt{1 - (\frac{p_3}{p_1})^2} &
0 & -\frac{p_3}{p_1} \end{array} \right) \\
&& \cdot \left( \begin{array}{ccc} 1 & 0 & 0 \\
0 & \cos (g \, q_3) & \sin (g \, q_3) \\
 0 & -\sin (g \, q_3) & \cos (g \, q_3) \end{array} \right).
\end{eqnarray*}

The transformation equations can also be obtained from a generating
functional of the form
\begin{equation} \label{gefu}
F\bigl[ p_1, q_2, p_3, \{ Q_k^{a'} \}, \{ \Pi_{ka} \} \bigr] = \int \!
{\cal F}({\bf x}) \, d^3 {\bf x},
\end{equation}
where
\begin{eqnarray*}
{\cal F} &\!\! = \!\!& \frac{2}{g} \, \eta \, p_1 \arctan \left(
\frac{\sqrt{1 - (\frac{p_3}{p_1})^2} \, || \Pi_1 || - \left[ \Pi_{11}
\sin (g \, q_2) + \Pi_{12} \cos (g \, q_2) \right]}{\sqrt{\Pi_{13}^2 -
(\frac{p_3}{p_1})^2 \, || \Pi_1 ||^2 + \left[ \Pi_{11} \cos (g \, q_2)
- \Pi_{12} \sin (g \, q_2)\right]^2}} \right) \\* && 
+ \, \frac{2}{g} \, p_3 \arctan \Biggl( \biggl[ || \Pi_1 || \, \left[
N_1 \sin (g \, q_2) + N_2 \cos (g \, q_2) \right] \sqrt{1 -
(p_3/p_1)^2} \\* 
&& \hspace{7.0em} + \, \eta \, ||N|| \, \Bigl[ \Pi_{13}^2 - \Bigl(
\frac{p_3}{p_1} \Bigr)^2 \, || \Pi_1 ||^2 + \left[ \Pi_{11} \cos (g \,
q_2) - \Pi_{12} \sin (g \, q_2) \right]^2 \Bigr]^{1/2} \biggr]
\\[1.0ex] 
&& \hspace{5.7em} \times \Bigl[ \Bigl( \sqrt{1 - (p_3/p_1)^2} \, K_1 -
\frac{p_3}{p_1} ||N|| \, \Pi_{11} \Bigr) \sin (g \, q_2) \\* 
&& \hspace{7.0em} + \Bigl( \sqrt{1 - (p_3/p_1)^2} \, K_2 -
\frac{p_3}{p_1} ||N|| \, \Pi_{12} \Bigr) \cos (g \, q_2) \Bigr] \bigg/
\\ 
&& \hspace{6.5em} \biggl[ - (K_1^2 + K_2^2) + \Bigl( \frac{p_3}{p_1}
\Bigr)^2 || \Pi_1 ||^2 N_3^2 + \left[ K_1 \cos (g \, q_2) - K_2 \sin
(g \, q_2) \right]^2 \\* 
&& \hspace{7.0em} + \Bigl( \frac{p_3}{p_1} \Bigr)^2 || \Pi_1 ||^2
\left[ N_1 \cos (g \, q_2) - N_2 \sin (g \, q_2) \right]^2 \biggr]
\Biggr) \\[1.0ex] 
&& - Q^{k a'} \, {\Omega_{a'}}^b \, \Pi_{kb} + \frac{1}{2 g} \,
{\varepsilon_{bc}}^a \left( \Omega^T \, \partial_3 \, \Omega
\right)^{cb} \Pi_{3a} - \frac{1}{2 g} \Pi_{11} N_1 \, \partial_2
\left( \frac{1}{\Pi_{12}^2 + \Pi_{13}^2} \right) \\[1.5ex] 
&& + \frac{1}{g} \, \frac{\Pi_{21} (\Pi_{13} \, \partial_2 \Pi_{12} -
\Pi_{12} \, \partial_2 \Pi_{13}) + \Pi_{11} (\Pi_{13} \, \partial_1
\Pi_{12} - \Pi_{12} \, \partial_1 \Pi_{13})}{\Pi_{12}^2 + \Pi_{13}^2}
\\[1.5ex] 
&& - \frac{1}{g} \left[ \partial_1 \, || \Pi_1 || + \partial_2 \left(
|| \Pi_1 || \, \frac{\Pi_{12} \Pi_{22} + \Pi_{13} \Pi_{23}}{\Pi_{12}^2
+ \Pi_{13}^2} \right) \right] \arctan \left( \frac{K_1}{|| \Pi_1 ||
N_1} \right) \\[1.5ex] 
&& + \frac{1}{2 g} \left[ \partial_2 \left( \frac{\Pi_{11}
N_1}{\Pi_{12}^2 + \Pi_{13}^2} \right) \right] \log \left(
\frac{||N||^2}{M^8}\right)
\end{eqnarray*}
and
$$ K_a = {\varepsilon_a}^{bc} \, \Pi_{1b} N_c = \left( {\delta_a}^c
\delta^{bd} - {\delta_a}^d \delta^{bc} \right) \Pi_{1b} \Pi_{1c}
\Pi_{2d}. $$ 
The components $N_a$ are those defined in equation (\ref{tep}) and $M$
denotes a constant with the dimension of energy. There is also a real
phase $\eta$ which can take the values $\pm 1$. Expression (\ref{tep})
is used for the matrix $\Omega$, and the primed index $a'$ stands as a
reminder of the fact that only independent components $Q_k^{a'}$,
i.e. those included in the list (\ref{finvar}) should be summed
over. Now the transformation equations read
\begin{subequations}
\begin{eqnarray}
q_1({\bf x}) &=& \frac{\delta F}{\delta p_1({\bf x})} \label{gfeq1}
\\* 
p_2({\bf x}) &=& - \, \frac{\delta F}{\delta q_2({\bf x})}
\label{gfeq2} \\ 
q_3({\bf x}) &=& \frac{\delta F}{\delta p_3({\bf x})} \label{gfeq3} \\ 
P_{ka'}({\bf x}) &=& - \, \frac{\delta F}{\delta Q^{ka'}({\bf x})}
\label{gfeq4} \\* 
A_k^a({\bf x}) &=& - \, \frac{\delta F}{\delta \Pi_a^k({\bf x})}.
\label{gfeq5}
\end{eqnarray}
\end{subequations}
Equations (\ref{gfeq1}) -- (\ref{gfeq3}) reproduce equations
(\ref{ftq1}), (\ref{ftp2}) and (\ref{ftq3}) in a form where the
components $G_a$ are expressed in the variables $\{ p_1, q_2, p_3,
\Pi_{1a} \}$ by inverting equations (\ref{ftp1}), (\ref{ftq2}) and
(\ref{ftp3}), i.e.,
\begin{eqnarray} \label{gegen}
G_1 &\! = \!& \Biggl( \frac{p_3}{p_1} \, || \Pi_1 || \, \left[
\Pi_{11} \cos (g \, q_2) - \Pi_{12} \sin (g \, q_2) \right] \nonumber
\\* 
&& \hspace{1.0em} - \eta \, \Pi_{13} \sqrt{\Pi_{13}^2 - \left(
\frac{p_3}{p_1} \right)^2 || \Pi_1 ||^2 + \left[ \Pi_{11} \cos (g \,
q_2) - \Pi_{12} \sin (g \, q_2)\right]^2} \, \Biggr) \nonumber \\ 
&& \times \, \frac{p_1 \cos (g \, q_2)}{\Pi_{13}^2 + \left[ \Pi_{11}
\cos (g \, q_2) - \Pi_{12} \sin (g \, q_2)\right]^2} \nonumber
\\[1.0ex] 
G_2 &\! = \!& - \Biggl( \frac{p_3}{p_1} \, || \Pi_1 || \, \left[
\Pi_{11} \cos (g \, q_2) - \Pi_{12} \sin (g \, q_2) \right] \nonumber
\\* 
&& \hspace{1.8em} - \eta \, \Pi_{13} \sqrt{\Pi_{13}^2 - \left(
\frac{p_3}{p_1} \right)^2 || \Pi_1 ||^2 + \left[ \Pi_{11} \cos (g \,
q_2) - \Pi_{12} \sin (g \, q_2)\right]^2} \, \Biggr) \nonumber \\ 
&& \times \, \frac{p_1 \sin (g \, q_2)}{\Pi_{13}^2 + \left[ \Pi_{11}
\cos (g \, q_2) - \Pi_{12} \sin (g \, q_2)\right]^2} \\[1.0ex] G_3 &\!
= \!& \Biggl( \frac{p_3}{p_1} \, || \Pi_1 || \, \Pi_{13} + \eta \left[
\Pi_{11} \cos (g \, q_2) - \Pi_{12} \sin (g \, q_2) \right] \nonumber
\\* 
&& \hspace{1.0em} \times \sqrt{\Pi_{13}^2 - \left( \frac{p_3}{p_1}
\right)^2 || \Pi_1 ||^2 + \left[ \Pi_{11} \cos (g \, q_2) - \Pi_{12}
\sin (g \, q_2)\right]^2} \, \Biggr) \nonumber \\ 
&& \times \, \frac{p_1}{\Pi_{13}^2 + \left[ \Pi_{11} \cos (g \, q_2) -
\Pi_{12} \sin (g \, q_2)\right]^2}, \nonumber
\end{eqnarray}
where the sign $\eta$ must be chosen so that
\begin{eqnarray*}
&& \hspace{-1.0em} \Biggl( \frac{p_3}{p_1} \, || \Pi_1 || \, \left[
\Pi_{11} \cos (g \, q_2) - \Pi_{12} \sin (g \, q_2) \right] \\* 
&& \hspace{-0.0em} - \eta \, \Pi_{13} \sqrt{\Pi_{13}^2 - \left(
\frac{p_3}{p_1} \right)^2 || \Pi_1 ||^2 + \left[ \Pi_{11} \cos (g \,
q_2) - \Pi_{12} \sin (g \, q_2)\right]^2} \, \Biggr) \geq 0.
\end{eqnarray*}
Equations (\ref{gfeq4}) and (\ref{ftP}) are equivalent, and with the
help of equations (\ref{ftq3}), (\ref{ftP}) and (\ref{gegen}) it is
also possible to see the equivalence of equations (\ref{gfeq5}) and
(\ref{finintra}). Although the generating functional looks rather
complicated, its mere existence is sufficient to confirm that the
transformation (\ref{fintrans}) is canonical. We have now all the
necessary tools at hand for constructing the physical Hamiltonian.

\section{Physical variables}

The greatest advantage in passing to the new variables (\ref{finvar})
is the fact that their behaviour in the limit $G_a \rightarrow 0$ is
relatively simple to analyse. Of course, if we were to be exact, we
would have to specify this limit precisely by starting from equation
(\ref{gdef}) and then defining suitable norms and function spaces for
the fields $A_k^a$ and $\Pi_{ka}$.  Instead of doing so I will adopt a
physicist's point of view and assume that it does not matter much
which particular function spaces we use if our fields are sufficiently
smooth and vanish rapidly enough at infinity. Looking at equations
(\ref{ftp1}) -- (\ref{ftp3}) we see then that Gauss's law is
implemented in the new variables by setting
\begin{equation} \label{glaw}
p_1 = p_2 = p_3 = 0.
\end{equation}
That these constraints are preserved in time in the dynamics described
by the Hamiltonian (\ref{ham}) is evident because $p_1$ and $p_2$ are
constants of motion and $\dot{p_3}$ is proportional to the Gauss law
generators. Equations (\ref{ftq1}) -- (\ref{ftq3}) reveal similarly
that the angles $q_1$, $q_2$ and $q_3$ become ambiguous when $G_a
\rightarrow 0$ and therefore we must discard these variables as
nonphysical. The physical variables are then the pairs $(Q_k^{a'},
P_{ka'})$, as their defining equations (\ref{ftQ}) and (\ref{ftP}) are
independent of $G_a$.  Since the generating functional (\ref{gefu})
does not contain explicit time dependence, the dynamics of $Q_k^{a'}$
and $P_{ka'}$ is governed by the Hamiltonian (\ref{ham}) under the
constraint (\ref{glaw}), i.e.,
$$ H_{\textrm{phys}} = H_{| \, p_1 \, = \, p_2 \, = \, p_3 \, = \,
  0}. $$ 
A formula for the Hamiltonian (\ref{ham}) in the variables
(\ref{finvar}) is most easily obtained with the help of equations
(\ref{finintr}). Since the Hamiltonian is invariant under gauge
transformations of this form, we immediately get the result
$$ H = \int \! \left( \frac{1}{2} \, P_{ka} P^{ka} + \frac{1}{4} \,
\widetilde{\Phi}_{kl}^a \, \widetilde{\Phi}^{kl}_a \right) d^3 {\bf
x}, $$ 
where
\begin{eqnarray*}
\widetilde{\Phi}_{kl}^a &=& \partial_l \widetilde{Q}_k^a - \partial_k
\widetilde{Q}_l^a + g \, {\varepsilon_{bc}}^a \widetilde{Q}_k^b \,
\widetilde{Q}_l^c, \\*[1.0ex] 
\widetilde{Q}_k^a &=& Q_k^a + \frac{1}{g} \, \delta_{k1} \,
{(O_4)^a}_2 \frac{p_1}{P_{11}} \sqrt{1 - \left( \frac{p_3}{p_1}
\right)^2} \, - \, \frac{1}{g} \, \delta_{k2} \, {\delta^a}_2
\frac{p_3}{P_{23}}
\end{eqnarray*}
and the definitions (\ref{konsta}) are implied. Imposing the
constraint (\ref{glaw}) it is then easy to see that
\begin{equation} \label{pham}
H_{\textrm{phys}} = \int \! \left( \frac{1}{2} \, P_{ka} P^{ka} +
\frac{1}{4} \, \Phi_{kl}^a \, \Phi^{kl}_a \right) d^3 {\bf x},
\end{equation}
where
$$ \Phi_{kl}^a = \partial_l Q_k^a - \partial_k Q_l^a + g \,
{\varepsilon_{bc}}^a Q_k^b \, Q_l^c $$ 
and
\begin{eqnarray} \label{komis}
Q_1^2 &=& - \, \frac{1}{g \, P_{11}} \sum_{k=2}^3 \left( \partial_k \,
P_{k3} - g \, {{\varepsilon_b}^c}_3 Q_k^b \, P_{kc} \right) \nonumber
\\* 
Q_1^3 &=& \frac{1}{g \, P_{11}} \Bigl( \partial_3 \, P_{32} -
\sum_{k=2}^3 g \, {{\varepsilon_b}^c}_2 Q_k^b \, P_{kc} \Bigr) \\
Q_2^2 &=& \frac{1}{g \, P_{23}} \left( \partial^{\, k} \, P_{k1} - g
\, {{\varepsilon_b}^c}_1 \, Q_3^b \, P_{3c} \right) \nonumber
\\*[2.0ex] 
P_{12} &=& P_{13} \hspace{0.3em} = \hspace{0.3em} P_{22}
\hspace{0.3em} = \hspace{0.3em} 0. \nonumber
\end{eqnarray}
Equations (\ref{ftP}) and (\ref{tep}) also show that
$$ P_{11} \geq 0, \qquad P_{23} \geq 0. $$
It may be a little surprising that the Hamiltonian (\ref{pham}) is
local, because one would expect the Gauss law to produce nonlocal
terms. However, the locality of $H_{\textrm{phys}}$ becomes easy to
understand if we look at the definitions (\ref{komis}). Our gauge
choices have annihilated three momentum components, and when we solve
Gauss's law for the coordinates conjugate to these momenta, the result
is local. One should also note that the Hamiltonian density is
singular at points where $P_{11}$ or $P_{23}$ vanishes.  These are
exactly the same points where the gauge transformation matrix
(\ref{tep}) becomes ambiguous.

Now we would like to examine what the Hamiltonian (\ref{pham}) looks
like at small and large values of the coupling constant $g$.  For that
purpose we note that every component $Q_k^a$ consists of terms
proportional to $g^{-1}$ and terms independent of $g$.  Therefore the
field tensor components $\Phi_{kl}^a$ range from $g^{-1}$ to $g^1$ and
as a result, the Hamiltonian density takes the form
\begin{equation} \label{komposti}
{\cal H}_{\textrm{phys}} = \frac{1}{2 g^2} {\cal H}^{(0)} + \,
\frac{1}{g} \, {\cal H}^{(1)} + \, {\cal H}^{(2)} + g \, {\cal
H}^{(3)} + \frac{g^2}{2} {\cal H}^{(4)}.
\end{equation}
At small values of $g$ the dominant term is ${\cal H}^{(0)}$, and it
is rather straightforward to work out that
\begin{eqnarray} \label{ham0}
{\cal H}^{(0)} &\hspace{-0.6em} = \hspace{-0.3em}& \hspace{-0.3em}
\left[ - \partial_2 \left( \frac{1}{P_{11}} \sum_{k=2}^3 \partial_k \,
P_{k3} + \frac{P_{21}}{P_{11} P_{23}} \sum_{k=1}^3 \partial_k \,
P_{k1} \right) - \partial_1 \left( \frac{1}{P_{23}} \sum_{k=1}^3
\partial_k \, P_{k1} \right) \right]^2 \nonumber \\* 
&& \hspace{-0.3em} + \left[ - \partial_3 \left( \frac{1}{P_{11}}
\sum_{k=2}^3 \partial_k \, P_{k3} + \frac{P_{21}}{P_{11} P_{23}}
\sum_{k=1}^3 \partial_k \, P_{k1} \right) \right]^2 + \left[
\partial_2 \left( \frac{\partial_3 P_{32}}{P_{11}} \right) \right]^2
\\ 
&& \hspace{-0.3em} + \left[ \partial_3 \left( \frac{\partial_3
P_{32}}{P_{11}} \right) \right]^2 + \left[ - \, \frac{\partial_3
P_{32}}{P_{11} P_{23}} \sum_{k=1}^3 \partial_k \, P_{k1} \right]^2 +
\left[ \partial_3 \left( \frac{1}{P_{23}} \sum_{k=1}^3 \partial_k \,
P_{k1} \right) \right]^2. \nonumber
\end{eqnarray}
This expression looks a bit complicated, but it is noteworthy that
${\cal H}^{(0)}$ does not depend on the coordinates $Q_k^a$. At large
values of $g$ we similarly find the dominant term to be
\begin{eqnarray} \label{ham4}
{\cal H}^{(4)} &=& \Biggl\{ \Bigl[ \bigl( P_{33} Q_3^1 - P_{31} Q_3^3
\bigr) \bigl( P_{33} Q_3^2 - P_{32} Q_3^3 \bigr) + P_{23} Q_3^2 \bigl(
P_{33} Q_2^1 - P_{31} Q_2^3 \bigr) \nonumber \\*[-1.0ex] 
&& \hspace{1.5em} + P_{23} P_{32} \bigl( Q_2^3 Q_3^1 - Q_2^1 Q_3^3
\bigr) \Bigr]^2 \nonumber \\[0.5ex] 
&& \hspace{0.9em} + P_{23}^2 \Bigl[ - P_{11} Q_1^1 Q_2^3 + Q_2^1
\bigl( P_{23} Q_2^1 - P_{21} Q_2^3 + P_{33} Q_3^1 - P_{31} Q_3^3
\bigr) \Bigr]^2 \nonumber \\[1.5ex] 
&& \hspace{0.9em} + \Bigl[ P_{23} Q_2^1 \bigl( - P_{32} Q_3^1 + P_{31}
Q_3^2 \bigr) - \bigl( P_{11} Q_1^1 + P_{21} Q_2^1 \bigr) \bigl( P_{33}
Q_3^2 - P_{32} Q_3^3 \bigr) \Bigr]^2 \nonumber \\[1.5ex] 
&& \hspace{0.9em} + \Bigr[ - P_{23} Q_3^2 \bigl( P_{23} Q_2^1 - P_{21}
Q_2^3 + P_{33} Q_3^1 \bigr) \nonumber \\* 
&& \hspace{2.4em} + Q_3^3 \bigl( P_{23} P_{32} Q_3^1 + P_{21} P_{33}
Q_3^2 \bigr) - P_{21} P_{32} \bigl( Q_3^3 \bigr)^2 \Bigr]^2 \\[1.0ex]
&& \hspace{0.9em} + P_{23}^2 \Bigl[ Q_3^1 \bigl( P_{23} Q_2^1 - P_{21}
Q_2^3 + P_{33} Q_3^1 \bigr) - Q_3^3 \bigl( P_{11} Q_1^1 + P_{31} Q_3^1
\bigr) \Bigr]^2 \nonumber \\[1.5ex] 
&& \hspace{0.9em} + \Bigr[ - P_{23} P_{32} \bigl( Q_3^1 \bigr)^2 +
P_{23} Q_3^2 \bigl( P_{11} Q_1^1 + P_{31} Q_3^1 \bigr) \nonumber \\*
&& \hspace{2.4em} + P_{21} Q_3^1 \bigl( - P_{33} Q_3^2 + P_{32} Q_3^3
\bigr) \Bigr]^2 \nonumber \\[1.0ex] 
&& \hspace{0.9em} + P_{11}^2 \Bigl[ - P_{23} Q_2^3 Q_3^2 - Q_3^3
\bigl( P_{33} Q_3^2 - P_{32} Q_3^3 \bigr) \Bigr]^2 \nonumber \\[1.0ex]
&& \hspace{0.9em} + \bigl( P_{11} P_{23} \bigr)^2 \bigl( Q_2^3 Q_3^1 -
Q_2^1 Q_3^3 \bigr)^2 \nonumber \\[0.5ex] 
&& \hspace{0.9em} + P_{11}^2 \Bigl[ P_{23} Q_2^1 Q_3^2 + Q_3^1 \bigl(
P_{33} Q_3^2 - P_{32} Q_3^3 \bigr) \Bigr]^2 \Biggr\} \bigg/ \bigl(
P_{11} P_{23} \bigr)^2. \nonumber
\end{eqnarray}
This is also a rather complicated expression, being fourth order in
the coordinates $Q_k^a$ and fractional in the momenta $P_{ka}$.
Finally we should note that the form of the decomposition
(\ref{komposti}) is actually a matter of choice, since it is always
possible to scale the variables by
$$ (Q_k^a, P_{ka}) \rightarrow (g^{\alpha_{ka}} Q_k^a, \,
g^{-\alpha_{ka}} P_{ka}), $$ 
where the $\alpha_{ka}$'s are arbitrary constants. However, scalings
like this would alter the $g$-dependence of the field tensor
$\Phi_{kl}^a$ and the covariant derivative. As a result, the
interpretation of $g$ would also change. In the present form $g$ is
defined so that the limit $g \rightarrow 0$ corresponds to an Abelian
theory. The singular behaviour of ${\cal H}_{\textrm{phys}}$ in this
limit then stems from an obvious qualitative difference between
Abelian and non-Abelian theories in the function group method. Namely,
the algebra (\ref{alg}) shows that for a non-Abelian theory $(g \neq
0)$ the Gauss law generators must be parametrised with variables that
contain one canonically conjugate pair, whereas in the Abelian case
$(g = 0)$ this parametrisation cannot contain canonical pairs at
all. Also the solution of Gauss's law given in (\ref{komis}) is
genuinely non-Abelian and impossible to extend to the Abelian case.

When quantising the Hamiltonian (\ref{pham}), we could try to quantise
one of the limiting cases (\ref{ham0}) or (\ref{ham4}) first and then
develop a perturbation expansion in appropriate powers of $g$. At
least the weak coupling Hamiltonian (\ref{ham0}), despite its
complicated appearance, looks easy to quantise as its eigenstates
would consist of common eigenstates of the momentum operators. The
strong coupling Hamiltonian (\ref{ham4}) is considerably more
difficult to quantise in the canonical approach because we would have
to solve problems connected with the ordering of operators and with
the regularisation of higher order functional derivatives defined at
the same point in space. Moreover, it is not clear whether large
values of the bare coupling constant are physically relevant. As a
general feature of quantisation one should also take into account that
two classical systems connected by a canonical transformation do not
necessarily yield unitarily equivalent quantum systems. For example,
in quantum mechanics it is often difficult to find a unitary
transformation corresponding to action-angle variables in classical
mechanics \cite{ms}. The fact that the transformation (\ref{fintrans})
is nonlinear might thus have an effect on the quantisation of the
Hamiltonian (\ref{pham}). Finding a suitable quantisation procedure
remains a problem to study.

\section{Conclusions}

The unconstrained Hamiltonian (\ref{pham}) lies at the end of a long
journey which started from the temporal gauge Hamiltonian (\ref{ham})
and passed through the transformation (\ref{fintrans}), making it
finally possible to implement Gauss's law in the new variables
(\ref{finvar}). The canonical pairs $(q_i, p_i)$ turned out to be
nonphysical, which led to the conclusion that the physical degrees of
freedom are described by the gauge-invariant fields $(Q_k^{a'},
P_{ka'})$. Equation (\ref{komis}) then defined those components that
are not free variables. The actual construction of the variables
(\ref{finvar}) relied on the parametrisation (\ref{trans}) and the
complementary choices (\ref{q1def}), (\ref{p3def}) and
(\ref{q3def}). One could also easily experiment with different choices
and derive alternative Hamiltonians corresponding to them by applying
the general principles stated in section \ref{konstru}. In particular,
the Poisson bracket relations (\ref{pbr1}) allow for a large variety
of possible U(1) gauges, given only that the initial choice
(\ref{intrans}) is made.

An extension of this construction to more general Lie groups is
relatively straightforward to outline. One should begin by deriving a
parametrisation of the Gauss law generators similar to equation
(\ref{trans}). Identifying the $G_a$'s with elements of the
corresponding Lie algebra, one should select the maximum number of new
canonical momenta from the maximal Abelian subspace of the enveloping
algebra. The remaining variables needed for the parametrisation should
then be chosen so that the Lie algebra relations
$$ \{ G_a({\bf x}), G_b({\bf y}) \} = - g \, {f_{ab}}^c \, G_c({\bf
y}) \delta({\bf x} - {\bf y}) $$ 
would hold as a consequence of the canonical Poisson brackets. After
identifying which variables are gauge-dependent one should define
gauge-invariant variables by transforming the gauge-dependent degrees
of freedom away. The details of the construction would then depend on
the Poisson brackets of the gauge-invariant variables and the way of
defining those gauge degrees of freedom that are not fixed by the
parametrisation of the Gauss law generators. No doubt that the
calculations would be much more complicated than in the SU(2) case. In
addition to generalising the Lie group, one could also extend the
method by adding matter fields, in particular fermions, into the
theory. However, it seems that the question of quantisation deserves
the most attention in the future because it is crucial for the
physical applicability of the Hamiltonian (\ref{pham}).

\begin{acknowledgements}
I am indebted to professor C. Cronstr\"{o}m for suggesting this
problem and for numerous discussions and comments. This work was
partially supported by the Magnus Ehrnrooth foundation and the Finnish
Cultural Foundation.
\end{acknowledgements}

\pagebreak

\end{document}